# Fast Extended Depth of Focus Meta-Optics for Varifocal Functionality

JAMES E. M. WHITEHEAD,[1] ALAN ZHAN,[3] SHANE COLBURN,[3] LUOCHENG HUANG,[1] AND ARKA MAJUMDAR[1, 2, *]

[1]*Department of Electrical and Computer Engineering, University of Washington, Seattle, Washington 98195, USA.*
[2]*Department of Physics, University of Washington, Seattle, Washington 98195, USA.*
[3]*Tunoptix, 4000 Mason Road #300, Fluke Hall, Seattle Washington 98195, USA.*
*[*arka@uw.edu](mailto:arka@uw.edu)*

**Abstract:** Extended depth of focus (EDOF) optics can enable lower complexity optical imaging systems when compared to active focusing solutions. With existing EDOF optics, however, it is difficult to achieve high resolution and high collection efficiency simultaneously. The subwavelength pitch of meta-optics enables engineering very steep phase gradients, and thus meta-optics can achieve both a large physical aperture and high numerical aperture. Here, we demonstrate a fast ($f/1.75$) EDOF meta-optic operating at visible wavelengths, with an aperture of 2 mm and focal range from 3.5 mm to 14.5 mm (286 diopters to 69 diopters), which is a 250 × elongation of the depth of focus relative to a standard lens. Depth-independent performance is shown by imaging at a range of finite conjugates, with a minimum spatial resolution of $\sim 9.84 \mu m$ (50.8 cycles/mm). We also demonstrate operation of a directly integrated EDOF meta-optic camera module to evaluate imaging at multiple object distances, a functionality which would otherwise require a varifocal lens.

## 1. Introduction

Optical imaging systems operating at finite conjugates suffer from a limited depth of focus. This often necessitates complex refocusing mechanisms. While large scale optics can be refocused by manually translating individual elements, integrated applications require high-precision actuators and active feedback to modify the optics. Several active solutions are currently used to adjust the focal plane in integrated systems such as electro-wetting[1] and MEMS actuation[2], [3]. These solutions, however, have various drawbacks such as delicate control mechanisms, extra electrical circuitry, multiple acquisitions, and temperature sensitivity. Instead of dynamically changing the focal plane, a static optic with an extended focal zone can be used to image all objects in a desired depth covered by the focal range [4]. For conventional lenses, the depth of focus is inversely proportional to the square of the numerical aperture $\left(NA \sim \frac{D}{2f}\right)$ where $D$ is the diameter, and $f$ is the focal length of the lens. This leads to a trade-off between the depth of focus and resolution. The depth of focus of any lens can be increased by reducing the NA but that limits the achievable resolution of the optical system due to the diffraction limit. Additionally, for many practical applications, we need a sufficiently large diameter lens to ensure collection of enough photons to achieve an acceptable signal-to-noise ratio.

Extended depth of focus (EDOF) refractive lenses have been demonstrated in the past by exploiting wavefront coding[4]–[6]. With the need for miniaturizing imaging systems for emerging applications like autonomous navigation, smart home, and the Internet of Things, there is a growing trend of migrating from refractive elements to flat diffractive optics. Unsurprisingly, EDOF concepts have also been demonstrated in diffractive optics in the recent past. Specifically, exploiting inverse design, an extreme depth of field for an EDOF lens was reported using multi-level diffractive optics[7]. The reported 1.8 mm EDOF lens had a smallest focal length of 5 mm, with the maximum NA being 0.16.

Owing to their subwavelength pitch, meta-optics can support higher phase gradients and thus larger NA compared to multi-level diffractive optics. Meta-optics are subwavelength diffractive optics, which can guide all the light to the zeroth order and can provide a full $2\pi$ phase shift using a binary mask [8], [9], making them compatible with a single stage lithography process. EDOF meta-optics have previously been utilized to mitigate chromatic aberrations in meta-optics [10], [11]. Recently, inverse design techniques have also been used to create EDOF meta-optics[12], [13]; however, the utility of EDOF meta-optics for finite conjugate imaging at different object and image distances has not been demonstrated. In this work, we report a 2 mm diameter EDOF meta-optic with a maximum numerical aperture of 0.28 ($f/\# = 1.75$). The extended depth of focus is 11 mm (from 3.5 mm to 14.5 mm), and thus the optical power in this system can be changed from 69 diopters to 286 diopters. We note that for a standard hyperboloid lens with a focal length at the center of the above range, the depth focus is expected to be $\sim 4\lambda \left(\frac{f}{D}\right)^2 \approx 43 \mu m$ for green light. Thus, our EDOF meta-optic shows a 250 × elongation of the depth of focus compared to a standard lens. We also demonstrate imaging in multiple finite conjugate planes, where the required focal length changes from 3.65 mm to 7.55 mm and we achieved a resolution of $\sim 9.84 \mu m$ (50.8 cycles/mm) in the object plane. Finally, we report direct integration of the meta-optic into a commercial camera module and show imaging at different object planes. Thus, exploiting a static meta-optic and computational back end, we can implement some functionalities of a varifocal lens. Transferring such optical functionality into the software could foster a new era of "software-defined optics".

## 2. Device Design and Fabrication

There are several different types of EDOF meta-optics that have been reported in the literature before [11]. Here, we employ a cubic EDOF[10], with a phase profile

$$\phi(x,y) = \frac{2\pi}{\lambda}\left(\sqrt{x^2 + y^2 + f^2} - f\right) + \frac{\alpha}{R^3}(x^3 + y^3)$$

Here, $\lambda$ is the operating wavelength, $x$ and $y$ are the cartesian coordinates, $f$ is the nominal focal length of the lens, $R$ is the radius of the phase mask, and $\alpha$ determines the strength of the cubic component. For our meta-optic, we set $\alpha$ to be $100\pi$, $R$ is $1\ mm$, and $f$ is $5.6\ mm$. We used silicon nitride (SiN) pillars on quartz as the scatterers for the meta-optic (Fig. 1a). For the scatterers, we used square posts to ensure polarization-insensitive operation, large phase coverage for a given height as well as faster writing speeds using electron-beam lithography. We simulate the transmission characteristics of these square posts using Lumerical FDTD Solutions under a periodic boundary conditions to ensure we have high transmission while covering the whole $0 - 2\pi$ phase-shift (Fig. 1b). The meta-optic is then created by mapping the desired phase to the appropriate scatterer geometry under the local phase approximation[9], [14]. A cubic meta-optic does not produce a lens-like point spread function (PSF), and to extract the depth of focus, we calculate the correlation of the PSF along the optical axis[11] (Fig. 1e). We define the depth of focus as the range where the correlation value stays above 0.5. From the numerical analysis, we estimate the depth of focus to be 11 mm, with a minimum focal length of 3.5 mm. This makes the maximum numerical aperture of the lens 0.28. Figs. 1c and 1d show the simulated PSF of the cubic EDOF at the minimum and maximum focal distances.

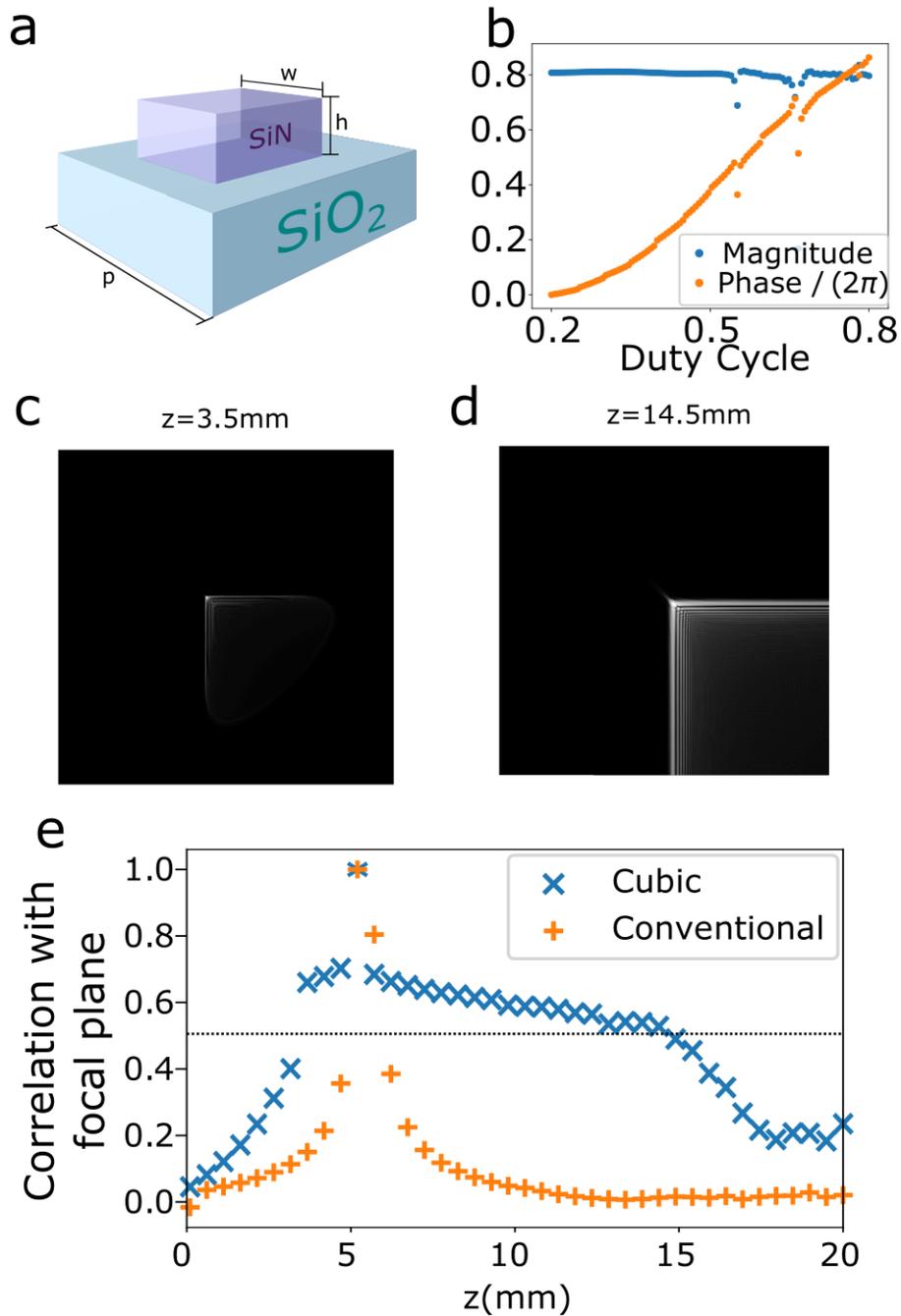

Fig. 1 Scatterer and meta-optics design and simulation: a) Schematic of $h = 633nm$ thick SiN square posts on a silicon oxide substrate. The periodicity p is kept constant, and the width w is changed to cover the whole $0 - 2\pi$ phase. b) Magnitude and phase of the transmitted light for a plane wave input with $p = 350nm$. c) and d) are simulated PSFs of the EDOF meta-optic at object distances of 3.5 mm and 14.5 mm respectively, e) Correlation plot of simulated PSF against the PSF at the central focal point for a cubic and a conventional meta-lens. The correlation clearly shows the extension of the depth of focus.

We used a 1.3 mm thick, 1 cm diameter silica wafer from Edmund Optics to fabricate the meta-optic. The choice of this wafer is motivated by its compatibility with commercial optical mounts. The double side polished quartz wafer was cleaned in a hot solution of sulfuric acid and hydrogen peroxide for 10 minutes. Plasma-enhanced chemical vapor deposition was then used to deposit a 633 nm layer of SiN on one side. A layer of 200 nm of ZEP-520A positive tone electron beam resist was spun after a short clean in oxygen plasma to maximize adhesion. An estimated 8 nm of Au/Pd was sputtered to dissipate charge produced from the electron beam. The pattern was then written using electron-beam lithography (JEOL JBX6300FS at 100 kV). An Au/Pd layer was then subsequently removed by immersing in transene gold etchant Type TFA with mild agitation. We then developed the resist in amyl acetate and rinsed in isopropyl alcohol, followed by a descum process in a weak oxygen plasma. A layer of 50 nm of nickel with a 5 nm chromium adhesion layer was then deposited using an electron-beam evaporator. This deposited Ni/Cr metal was selectively removed by dissolving the supporting resist in a solvent while sonicating. An inductively coupled fluorine plasma etcher (Oxford Plasmalab 100) etched the SiN layer down to the substrate and the Ni/Cr metal hard mask was removed with nickel and chromium etchants. The wafer was then cleaned using a hot solution of sulfuric acid and hydrogen peroxide for 10 minutes. Using direct write photolithography and electron-beam evaporation, a chromium aperture was patterned to block stray light from bypassing the meta-optics. Fig. 2 shows the optical and scanning electron microscopy (SEM) images of the fabricated EDOF meta-optic.

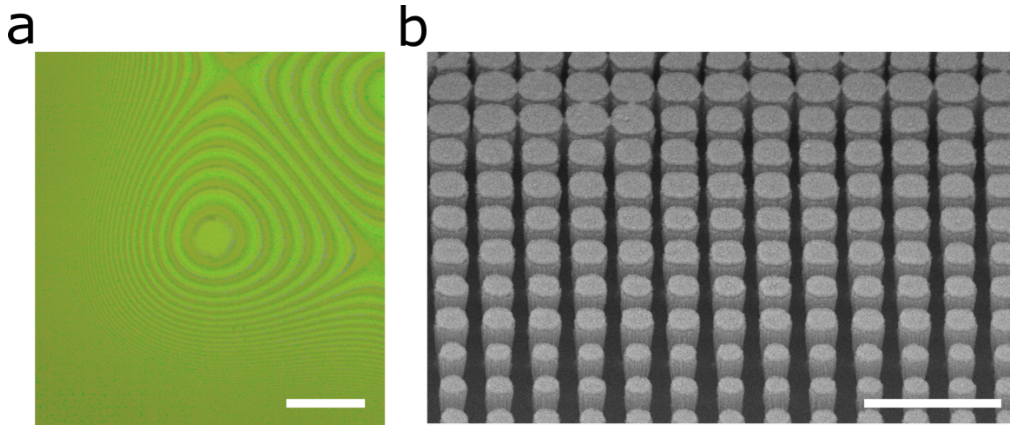

Fig. 2. Images of the fabricated meta-optics. The meta-optics was sputter coated with gold-palladium alloy to ensure charge dissipation a) Optical image (Scale bar is 150 $\mu m$). b) Scanning electron micrograph; scale bar of 1 $\mu m$ taken at $45^o$ to the normal.

### 3. Measurement

The meta-optic was then mounted in a 1cm diameter optic holder with standard C-mount threading for characterization. We first measured the PSF of the meta-optic: a 25 $\mu m$ diameter pinhole is used to approximate a point source, which is illuminated using a green (Thorlabs M530F1) light emitting diode (LED). The PSF is captured using a movable microscope (Fig. 3a). The position of the pinhole is changed from 7.3 mm to 15.1 mm away from the meta-optic, and the same range is used for the position of the camera (Fig. 3b). We can clearly observe that the PSF remains unchanged for various image and object distances. We note that, we are measuring the PSF for a finite conjugate system, as we will use a similar setup for imaging as well. We also confirmed that the minimum focal length for our meta-optic is ~3.5 mm, making the highest NA of the meta-optic 0.28, as predicted by the numerical simulation.

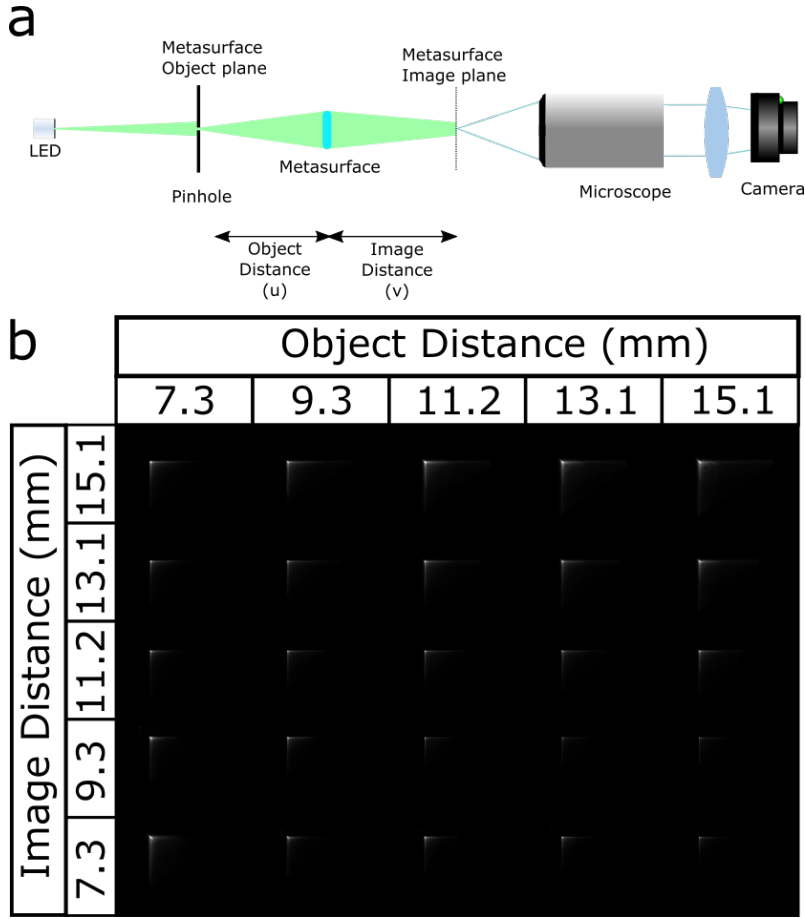

Fig 3. a) Setup for PSF measurement. b) Image of 25um pinhole. Image and object plane sweep. Illumination is using a 530nm LED with 33nm bandwidth.

We then tested our system by imaging a 1951 USAF Resolution Chart. The images of this backlit pattern were taken at various object ($u$) and image ($v$) distances to test the extended depth of focus capability of the lens. From ray optics, we can write for a lens with focal length $f$

$$\frac{1}{u} + \frac{1}{v} = \frac{1}{f}$$

As a cubic EDOF meta-optic does not produce a lens-like PSF and the raw captured images do not resemble the object, we need to deconvolve the captured sensor data to extract the in-focus image. Several different deconvolution routines, including Wiener, Richardson-Lucy, and learned methods can be used to extract the image[15], [16]. Here, we apply a routine based on Total Variation (TV) regularization to deconvolve and denoise the latent image (Fig. 4a). This deconvolution method optimizes the sum of the gradient magnitudes while deconvolving the image [17]. As we are imaging under incoherent illumination, we can model the whole imaging process as a linear system if the camera is not saturated. We ensured that the intensity of the measured PSF did not exceed the maximum threshold of the camera to conserve the linearity of the measured PSF. We also assumed that the PSF was shift-invariant. This assumption breaks down for large lateral displacements of the pinhole and taking multiple shifted PSFs while using a shift-variant deconvolution technique may produce better results[18]. In our work, this means we have a constrained field of view to ensure the shift-invariant property of our PSF. The imaging shows here show the same range of tunable focal lengths as from the PSF measurements. This achieved focus tuning, however, does not rely varifocal lens. We estimate the spatial resolution of our meta-optic via the linecuts from the measured patterns for an object (image) distance of 9.3 mm (13.1 mm) (Fig. 4b). The first column shows the linecuts from the horizontal lines and the vertical lines are shown in the second column. We can see that for the horizontal lines, we can differentiate lines of thickness of $\sim 9.84 \mu m$ (50.8 cycles/mm), whereas, for vertical lines only $\sim 11.05 \mu m$ lines (45.3 cycles/mm) can be resolved. We attribute this asymmetry to the asymmetric PSF of the cubic EDOF meta-optics. The diffraction limited resolution for a lens under the same condition to be $\sim 1.4 \mu m$. The lower spatial resolution comes from the extended PSF of the cubic meta-optic compared to a lens. As explained earlier, however, the EDOF meta-optic can be used to create images for different object and image distances, which is not possible with an ordinary metalens.

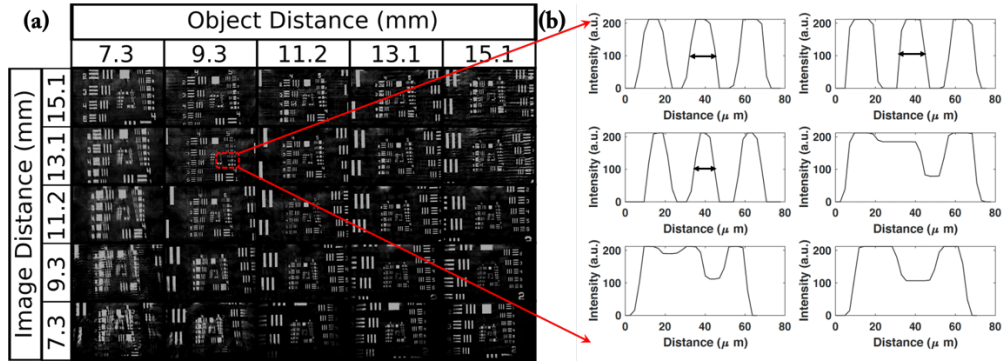

Figure 4. (a) Image of the Airforce Resolution chart for different image and object planes. The object distance is the separation between the transparency and the meta-optic while the image distance is the distance between the meta-optic and the camera. The object is illuminated via a 530nm LED with 33nm bandwidth (full width half maxima). (b) Line cut of the air force chart (for object distance of 9.3mm and image distance of 13.1mm) to estimate resolution: first column is for horizontal lines and second column is for vertical lines. Top to bottom row: group 5, number 4-6 for the Air Force Resolution chart.

Thus far in our experiments, we have used a microscope to relay the image produced by the EDOF meta-optic onto the sensor for imaging. With a larger aperture meta-optic, it is feasible to image directly onto an off-the-shelf camera module (E-consystems See3CAM_10CUG). For integration, the meta-optic is scribed into a circular piece, and mounted in a C-thread optic mount (Edmund Optics #63-979). The meta-optic is then attached to the camera module using a CS-mount, with a flange back distance of approximately 12 mm. The final package is shown in Fig. 5a. In this configuration, the imaging system has a full diagonal field of view of $28^o$.

We tested the imaging capabilities of the EDOF camera and compared it to an off-the-shelf plano-convex F/2 refractive singlet with focal length 6 mm (Edmund Optics #32-952) by imaging a series of QR codes at different object distances. Both the EDOF meta-optic and refractive singlet lens are tested using the same camera module and with the same CS-mount, ensuring the same sensor characteristics and magnification. The QR-codes are illuminated using a commercially available green LED ring light. The EDOF meta-optic and the refractive singlet have nominal focal lengths that are optimal for imaging at a finite conjugate distance of 13 mm. As seen in the Fig. 5b, both imaging systems are fully capable of resolving the QR code, but the refractive singlet displays improved brightness and contrast when compared to the EDOF meta-optic. As we move the object plane further out, however, the difference between the EDOF meta-optic and refractive singlet widens. For object distances of 25 mm, 50 mm, and 80 mm, the refractive singlet is incapable of resolving the QR code, while the EDOF meta-optic reliably resolves the barcodes throughout the range of object distances.

When compared to our EDOF meta-optic (F/2.8, focal length 5.6 mm), the refractive singlet has a similar focal length (6 mm vs 5.6 mm), but a larger diameter (3 mm vs 2 mm). The superior brightness and contrast of the singlet at the object distance of 13 mm is attributed both to the larger diameter of the refractive singlet, and also the properties of the EDOF meta-optic itself. At an aperture of 3 mm, the refractive optic has a surface that has twice the area of the EDOF meta-optic and is able to more than double the light collection. In addition, the EDOF meta-optic operates by sacrificing some contrast at the nominal object distance in favor of extending the depth-of-field significantly.

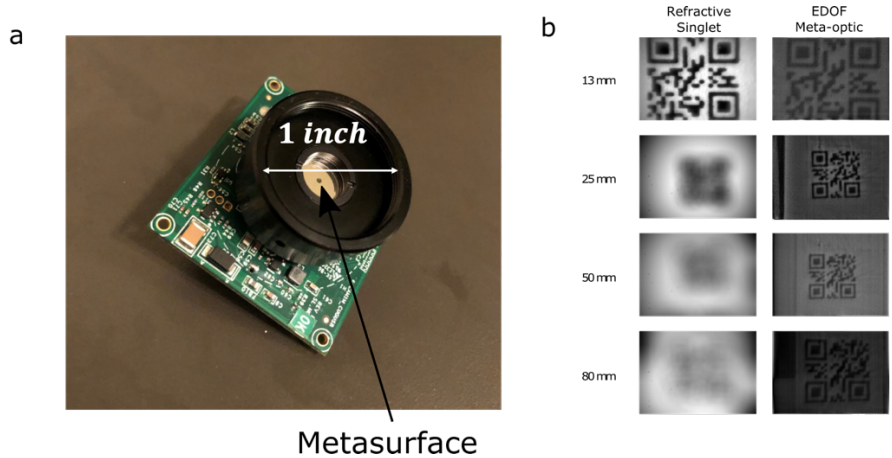

Fig. 5. a) Metasurface optic integrated with E-consystems camera module. b) Pictures of QR code object at differing object lengths taken by a singlet refractive lens with focal length 6mm at F/2 (left) and by a singlet EDOF meta-optic with nominal focal length 5.6mm at F/2.8 (right). The 13mm and 25mm object distances used a $5mm \times 5mm$ QR code. The 50mm object distance used a 10mm x 10mm QR code, and the 80mm object distance used a $30mm \times 30mm$ barcode.

## 4. Conclusion

We reported an EDOF, $f/1.75$ cubic meta-optic with a highest NA of 0.28, leveraging the subwavelength pitch of meta-optics to enable fast lensing. The EDOF nature of the meta-optic was evaluated using PSF measurements and imaging experiments. We also reported the development of an EDOF meta-optic camera module using a commercial sensor and demonstrated QR code imaging at different object distances. Our work opens opportunities for system-level integration of meta-optics with commercial sensors, and the simplification of tunable imaging systems exploiting EDOF properties. This new class of EDOF imaging systems could find applications in industrial production lines[19], compact image sensors, biological imaging[20], [21], automobile navigation, and driver monitoring systems. Going beyond EDOF, such software-defined optics could potentially transfer many of the hardware functionalities to software and accelerate the co-design and co-integration of hardware and software for free-space optics.

**Acknowledgement:** The research is supported by Tunoptix and DARPA (Contract no. 140D0420C0060). A.M. is also supported by a Washington Research Foundation distinguished investigator award. Part of this work was conducted at the Washington Nanofabrication Facility / Molecular Analysis Facility, a National Nanotechnology Coordinated Infrastructure (NNCI) site at the University of Washington with partial support from the National Science Foundation via awards NNCI-2025489 and NNCI-1542101.

**Competing Interests:** A.M., A.Z. and S.C. are involved with the startup Tunoptix Inc., which is commercializing technology discussed in this manuscript.

**References**

[1] S. W. Seo *et al.*, "Microelectromechanical-System-Based Variable-Focus Liquid Lens for Capsule Endoscopes," *Jpn. J. Appl. Phys.*, vol. 48, no. 5, p. 052404, May 2009, doi: 10.1143/jjap.48.052404.
[2] E. Arbabi, A. Arbabi, S. M. Kamali, Y. Horie, M. Faraji-Dana, and A. Faraon, "MEMS-tunable dielectric metasurface lens," *Nat. Commun.*, vol. 9, no. 1, p. 812, Feb. 2018, doi: 10.1038/s41467-018-03155-6.
[3] Z. Han, S. Colburn, A. Majumdar, and K. F. Böhringer, "MEMS-actuated metasurface Alvarez lens," *Microsyst. Nanoeng.*, vol. 6, no. 1, p. 79, Oct. 2020, doi: 10.1038/s41378-020-00190-6.
[4] E. R. Dowski and W. T. Cathey, "Extended depth of field through wave-front coding," *Appl. Opt.*, vol. 34, no. 11, pp. 1859–1866, Apr. 1995, doi: 10.1364/AO.34.001859.
[5] Zeev Zalevsky, "Extended depth of focus imaging: a review," *SPIE Rev.*, vol. 1, no. 1, pp. 1–11, Jan. 2010, doi: 10.1117/6.0000001.
[6] Rocha Karolinne Maia, "Extended Depth of Focus IOLs: The Next Chapter in Refractive Technology?," *J. Refract. Surg.*, vol. 33, no. 3, pp. 146–149, Mar. 2017, doi: 10.3928/1081597X-20170217-01.
[7] S. Banerji, M. Meem, A. Majumder, B. Sensale-Rodriguez, and R. Menon, "Extreme-depth-of-focus imaging with a flat lens," *Optica*, vol. 7, no. 3, pp. 214–217, Mar. 2020, doi: 10.1364/OPTICA.384164.
[8] S. M. Kamali, E. Arbabi, A. Arbabi, and A. Faraon, "A review of dielectric optical metasurfaces for wavefront control," *Nanophotonics*, vol. 7, no. 6, pp. 1041–1068, Jun. 2018, doi: https://doi.org/10.1515/nanoph-2017-0129.
[9] A. Zhan, S. Colburn, R. Trivedi, T. K. Fryett, C. M. Dodson, and A. Majumdar, "Low-Contrast Dielectric Metasurface Optics," *ACS Photonics*, vol. 3, no. 2, pp. 209–214, Feb. 2016, doi: 10.1021/acsphotonics.5b00660.
[10] S. Colburn, A. Zhan, and A. Majumdar, "Metasurface optics for full-color computational imaging," *Sci. Adv.*, vol. 4, no. 2, p. eaar2114, Feb. 2018, doi: 10.1126/sciadv.aar2114.
[11] L. Huang, J. Whitehead, S. Colburn, and A. Majumdar, "Design and analysis of extended depth of focus metalenses for achromatic computational imaging," *Photon Res*, vol. 8, no. 10, pp. 1613–1623, Oct. 2020, doi: 10.1364/PRJ.396839.
[12] E. Bayati, R. Pestourie, S. Colburn, Z. Lin, S. G. Johnson, and A. Majumdar, *Inverse Designed Extended Depth of Focus Meta-Optics for Broadband Imaging in the Visible*. 2021.
[13] E. Bayati, R. Pestourie, S. Colburn, Z. Lin, S. G. Johnson, and A. Majumdar, "Inverse Designed Metalenses with Extended Depth of Focus," *ACS Photonics*, vol. 7, no. 4, pp. 873–878, Apr. 2020, doi: 10.1021/acsphotonics.9b01703.
[14] R. Pestourie, C. Pérez-Arancibia, Z. Lin, W. Shin, F. Capasso, and S. G. Johnson, "Inverse design of large-area metasurfaces," *Opt. Express*, vol. 26, no. 26, pp. 33732–33747, Dec. 2018, doi: 10.1364/OE.26.033732.
[15] E. Tseng *et al.*, *Neural Nano-Optics for High-quality Thin Lens Imaging*. 2021.
[16] U. Akpinar, E. Sahin, M. Meem, R. Menon, and A. Gotchev, *Learning Wavefront Coding for Extended Depth of Field Imaging*. 2020.
[17] P. Getreuer, "Total Variation Deconvolution using Split Bregman," *Image Process. Line*, 2012, doi: 10.5201/ipol.2012.g-tvdc.
[18] E. Tseng *et al.*, *Neural Nano-Optics for High-quality Thin Lens Imaging*. 2021.

[19] "LIQUID LENSES IN MACHINE VISION." Edmund Optics. [Online]. Available: https://www.edmundoptics.com/ViewDocument/EO_Liquid_Lenses_in_Machine_vision_0419_EN.pdf

[20] B. Forster, D. Van De Ville, J. Berent, D. Sage, and M. Unser, "Complex wavelets for extended depth-of-field: a new method for the fusion of multichannel microscopy images.," *Microsc. Res. Tech.*, vol. 65, no. 1–2, pp. 33–42, Sep. 2004, doi: 10.1002/jemt.20092.

[21] F. Aguet, D. Van De Ville, and M. Unser, "Model-based 2.5-d deconvolution for extended depth of field in brightfield microscopy.," *IEEE Trans. Image Process. Publ. IEEE Signal Process. Soc.*, vol. 17, no. 7, pp. 1144–1153, Jul. 2008, doi: 10.1109/TIP.2008.924393.